 \documentclass{emulateapj}

\shorttitle{SDSSJ1402+6321}
\shortauthors{Bolton et al.}

\usepackage{natbib}
\begin{document}
\bibliographystyle{apj}

\title{SDSSJ140228.22+632133.3: \\ A New Spectroscopically
Selected Gravitational Lens\altaffilmark{1,}\altaffilmark{2}}

\author{Adam S. Bolton\altaffilmark{3},
 Scott Burles\altaffilmark{3}
 L\'{e}on V. E. Koopmans\altaffilmark{4},
 Tommaso Treu\altaffilmark{5,}\altaffilmark{7},
 and Leonidas A. Moustakas\altaffilmark{6}}

\altaffiltext{1}{
Based on observations made with the NASA/ESA Hubble Space Telescope,
obtained at the Space Telescope Science Institute, which is operated
by the Association of Universities for Research in Astronomy, Inc.,
under NASA contract NAS 5-26555.  These observations are associated
with program \#10174.  Support for program \#10174 was provided by
NASA through a grant from the Space Telescope Science Institute, which
is operated by the Association of Universities for Research in
Astronomy, Inc., under NASA contract NAS 5-26555.}
\altaffiltext{2}{
Also based on observations obtained under program GN-2004A-Q-5 at the
Gemini Observatory, which is operated by the Association of
Universities for Research in Astronomy, Inc., under a cooperative
agreement with the NSF on behalf of the Gemini partnership: the
National Science Foundation (United States), the Particle Physics and
Astronomy Research Council (United Kingdom), the National Research
Council (Canada), CONICYT (Chile), the Australian Research Council
(Australia), CNPq (Brazil) and CONICET (Argentina).}
\altaffiltext{3}{Department of Physics and Center for Space Research,
 Massachusetts Institute of Technology, 77 Massachusetts Ave.,
 Cambridge, MA 02139, USA ({\tt bolton@mit.edu, burles@mit.edu})}
\altaffiltext{4}{Kapteyn Institute, P.O. Box 800, 9700AV Groningen,
 The Netherlands ({\tt koopmans@astro.rug.nl})}
\altaffiltext{5}{Department of Physics and Astronomy, UCLA,
 Box 951547, Knudsen Hall, Los Angeles, CA 90095, USA
 ({\tt ttreu@astro.ucla.edu})}
\altaffiltext{6}{Space Telescope Science Institute, 3700 San Martin Dr.,
 Baltimore, MD 21218, USA ({\tt leonidas@stsci.edu})}
\altaffiltext{7}{Hubble Fellow}

\begin{abstract}
We present Gemini integral-field
unit (IFU)
spectroscopy and {\sl Hubble
Space Telescope} (HST) F435W- and F814W-band images of a newly
discovered four-image gravitational lens, SDSSJ140228.22+632133.3
(hereafter SDSSJ1402). The
system was identified as one of 49 gravitational-lens candidates in
the
luminous red galaxy
sample
of
the Sloan Digital Sky Survey,
based on higher-redshift emission lines in the spectra of the
lower-redshift galaxies.
We are imaging the most promising lens candidates
with HST as
part of a Snapshot program
designed to expand
the sample of
known gravitational lenses amenable to detailed photometric, lensing
and dynamical studies; SDSSJ1402 was the first
of our targets to be observed
with
the
ACS-WFC on board HST.
The lens is a smooth
elliptical galaxy at a redshift of
$z_{\rm l}=0.2046 \pm 0.0001$ with a
Sloan $r$-band magnitude of $17.00 \pm 0.05$
and a stellar velocity dispersion of
267 $\pm$ 17 km s$^{-1}$, obtained from its SDSS spectrum.
Multiple
emission lines place the quadruply-imaged source at a redshift
of $z_{\rm s}=0.4814 \pm 0.0001$.
The best-fitting singular isothermal ellipsoid lens model
gives an Einstein radius $b=1\farcs35 \pm 0\farcs05$
(or $[4.9 \pm 0.2]h_{65}^{-1}$ kpc), corresponding
to a total mass of
$(30.9 \pm 2.3)\times 10^{10}h_{65}^{-1}$~${\rm M}_{\sun}$
within the critical curve. In combination with HST photometry this
gives a rest-frame $B$-band mass-to-light ratio of
$(8.1 \pm 0.7)h_{65}$ times solar within the
same region. The
lens
model predicts a
luminosity-weighted
stellar dispersion within the
$3 \arcsec$-diameter
SDSS aperture of $\sigma_*\approx 270$~ km s$^{-1}$, in
good agreement with the observed value.
Using the model to {\sl de-lens} the four lensed images
yields
a source with a smooth, monotonically-decreasing
brightness distribution.
Taken in combination, the HST ACS images,
Gemini IFU spectroscopy, and self-consistent mass model
show SDSSJ1402 to be a genuine lens system.
\end{abstract}


\section{Introduction}

In the last decade, galaxy-size gravitational lenses have become an
increasingly
important
tool
for the study of
cosmology and galaxy evolution.
The number of known lenses has now reached almost 100
(see the CASTLES web page at
{\tt http://cfa-www.harvard.edu/castles/}), and
subsets of lenses with suitable properties are now available for a
variety of applications:
the determination of the
cosmological parameters from lens statistics
\citep[e.g.][]{turner_lensstat,
kochanek_lensstat_96, chae_lensstat_03}, the measurement of the
Hubble Constant from lens time delays \citep[e.g.][]{kundic_h0,
schechter_1115_delay, koopmans_1608_h0}, and the study
of the mass distribution of E/S0 galaxies and their dark matter halos
outside the local Universe
\citep[e.g.][]{kochanek_dyn_halo_94, rusin_lensstat, tk04}.

Despite
the great
progress, the small
subsample sizes
of suitable
lenses is a major limitation. This is particularly true for the study of
the properties of lens galaxies (typically E/S0s).
Most currently known lenses have been discovered
as bright quasars that show
multiple images when observed at high spatial resolution.
For optical quasar lenses, this biases the sample
in favor of systems where the
lensed source outshines the lens galaxy.
Radio surveys such as the Cosmic Lens All-Sky
Survey \citep[CLASS;][]{class_myers, class_browne}
are less susceptible to this bias, but 
redshifts in radio lens systems can be difficult
to obtain, and the lensing galaxies are often faint.
This situation makes it very
difficult to obtain high-quality photometry, redshifts, and internal
kinematics of lens galaxies,
which
are essential ingredients
for
detailed
modeling. It is not a coincidence that 3/5 lenses analyzed by the Lenses
Structure and Dynamics Survey
(LSD; \citealt{kt02, kt03, tk02, tk03, tk04};
hereafter collectively KT)
so far have been
discovered serendipitously in Hubble Space Telescope (HST) images.

To overcome this limitation we have initiated
the Sloan Lens ACS (SLACS) Survey, a new survey for
gravitational lenses that exploits the Sloan Digital Sky Survey (SDSS)
archive and the angular resolution of HST to find and confirm new
gravitational lenses. First,
a large sample of SDSS early-type galaxy spectra is searched
for emission lines at a higher
redshift than that of the absorption features
(\citealt{bolton_speclens}; see also \citealt{warren_0047_96}).
By requiring
that the foreground galaxy
dominate
the absorption spectrum, this procedure
will tend
to select bright lenses
with faint background sources. The second step consists of
follow-up snapshot observations of
the best candidates with the Advanced Camera for Survey
(ACS) to confirm
the incidence of lensing
and provide the lens geometry
necessary to construct a mass model\footnote{Our survey is the
second HST snapshot lens survey.  For details of the first,
see \citet{maoz_snapshot}.}.
Each candidate is observed for 420 seconds in each of F435W and F814W.
The F435W filter is selected to optimize the detection of the lensed
source galaxy
(expected to be blue and starforming) while F814W is selected to
optimize the signal-to-noise ratio on the (red) lens galaxy. Color
information also provides important evidence in support of the lensing
hypothesis.
In this {\it Letter} we present the first newly
discovered lens, along with ACS images and Gemini GMOS-North
integral-field spectroscopy and a lens model,
and discuss its overall properties.
We assume that the Hubble constant, the matter density,
and the cosmological constant are
H$_0=65\,h_{65}$~km\,s$^{-1}$\,Mpc$^{-1}$ (with $h_{65}=1$),
$\Omega_{\rm m}=0.3$, and $\Omega_{\Lambda}=0.7$, respectively.
All spectroscopic wavelengths are given in vacuum values.

\section{Observations}

\subsection{SDSS}

The original spectroscopic observations of SDSSJ1402 were obtained on
SDSS plate 605, fiber 503, MJD 52353
(included in the public Data Release 2).
The latest public SDSS photometric values for the galaxy
are $(g, r, i) = (18.36, 17.00, 16.48)$, all $\pm 0.05$,
and it forms part of the volume-limited luminous
red galaxy (LRG) spectroscopic sample \citep{eisenstein_lrg}.
The {\tt specBS} one-dimensional
spectroscopic pipeline
(Schlegel et al., in preparation; see also
\citealt{bolton_speclens})
determines a redshift $z = 0.2046 \pm 0.0001$
and a velocity dispersion of
267 $\pm$ 17 km s$^{-1}$ for SDSSJ1402.  Though
the spectrum is unambiguously that of an early-type galaxy at the
above redshift, it also exhibits nebular line emission at a background
redshift of $0.4814 \pm 0.0001$
\citep[see Figure 2 of][]{bolton_speclens}.  Given
the measured stellar velocity dispersion and these redshifts and
assuming a singular isothermal sphere galaxy model, one na\"{i}vely
expects a strong-lensing region of radius $2\farcs2 \pm 0\farcs 3$
(twice the
Einstein radius) in the plane of the sky, centered on the foreground
galaxy.  Considering the $1\farcs5$ radius of an SDSS spectrograph
fiber aperture, the prior probability of the system being a strong
gravitational lens
is essentially unity if one neglects the effects
of seeing and fiber misalignment,
and it was therefore
targeted for follow-up observation. \\

\subsection{HST-ACS}
\label{acs}

\emph{Hubble Space Telescope} (HST) observations of SDSSJ1402 were
obtained on 2004 August 4 with the Advanced Camera for Surveys (ACS),
with one 420-second exposure in each of the filters F435W and F814W.
The 5-$\sigma$ limits are approximately
$B_{435}\approx22$\,mag\,arcsec$^{-2}$ and
$I_{814}\approx23$\,mag\,arcsec$^{-2}$ in 2$\times$2-pixel apertures.
The brief exposure times result in comic rays affecting $\sim1$--$2$\%
of all pixels.  These are very effectively flagged and cleaned using
the Laplacian edge-detection technique (``LACOSMIC'') of
\citet{van_dokkum_lacosmic}.
The cosmics-cleaned images are spatially transformed to an
orthogonal coordinate system by the MULTIDRIZZLE routine as
implemented on pyraf by STScI.  This allows for high-precision
astrometry and spatially-correct galaxy shapes.
In each band, a 2D model of the lensing galaxy is computed using
GALFIT \citep{peng_galfit}.  The fits are
done assuming a simple S\'{e}rsic
profile
parametrized by index
$n$ (where $n=1$ is an exponential, and $n=4$ is a
de~Vaucouleur profile), as well as assuming a constrained de~Vaucouleur
profile.  The latter
is especially useful for comparison with
fundamental plane results.
The S\'{e}rsic fits
are sufficiently accurate to produce high-quality residual images,
which are used to measure the positions of candidate lensed images.
These are
identified
both
by
their image configurations and
by
their
similarity in $B_{435}-I_{814}$ colors.

The constrained de~Vaucouleur fit to the
F814W galaxy image gives a model
magnitude\footnote{All HST magnitudes are corrected
for dust extinction using \citet*{sfd_dust} maps.}
of $I_{814} = 16.28 \pm 0.01$, an isophotal
axis ratio $q_{814} = 0.786 \pm 0.01$,
a position angle
$\theta_{814} = 71.0^{\circ} \pm 0.2^{\circ}$
East of North, and an intermediate-axis effective
(half-light) radius
of $R_{\rm e} = 2\farcs 80 \pm 0\farcs 01$.
Fixing the galaxy shape from these
F814W de~Vaucouleurs structural parameters
and fitting for a magnitude in the F435 image yields
$B_{435} = 19.00 \pm 0.01$.
The unconstrained S\'{e}rsic fit to the F814W data
has an index of $n = 5.93 \pm 0.03$, indicating
a high degree of central concentration.

\begin{figure}
\begin{center}
\scalebox{1.15}{\plotone{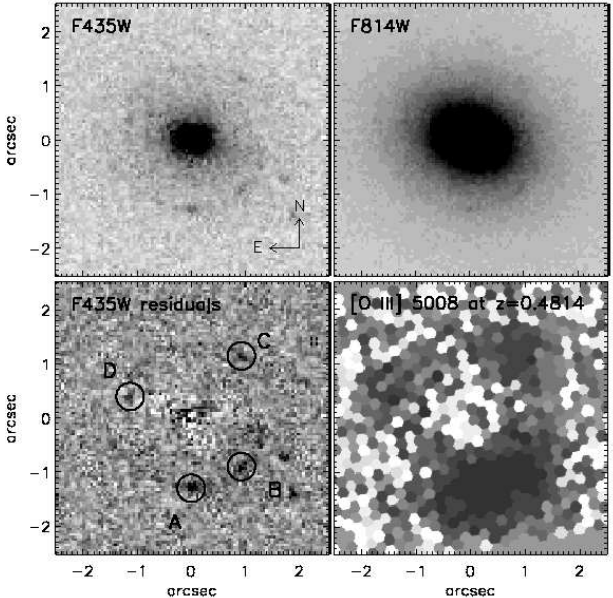}}
\end{center}
\caption{Imaging of SDSSJ1402.  Top panels:
ACS F435W (left) and ACS F814W (right) images.
Lower left: ACS F435W galaxy-fit-subtracted residual image,
with lensed image positions indicated.
Lower right: reconstructed 5-\AA-wide narrowband image
from GMOS-N IFU spectroscopy, centered at a
vacuum wavelength of 7420~\AA\, corresponding to redshifted [OIII]
5008.24~\AA.  Note coincidence of line emission with lensed image
positions. (IFU image scaling has been histogram-equalized.)
\label{ims}}
\end{figure}

Figure~\ref{ims} shows the ACS F435W and F814W images
and galaxy-subtracted F435W residual image of SDSSJ1402.
Despite low signal-to-noise, four images with similar
colors are easily identifiable in
both filters with a typical quad-like configuration;
we designate these images A--D.  Two of the
images (A and B) appear arc-like in both filters and stretched
tangentially.
We determine image centroids from the F435W image,
in which they are more pronounced and less
affected by residuals from the foreground galaxy
subtraction.  Table~\ref{imtable} lists
these image positions relative to the LRG
centroid.
We describe the determination of a gravitational-lens
model based on these image positions in
\S~\ref{lensmod} below.

\begin{table}
\centering
\begin{tabular}{cll}
  Image & $\Delta x$ (W) & $\Delta y$ (N) \\
  \hline
   A & $+$0\farcs00 & $-$1\farcs35 \\ 
   B & $+$0\farcs93 & $-$0\farcs95 \\ 
   C & $+$0\farcs88 & $+$1\farcs15 \\
   D & $-$1\farcs13 & $+$0\farcs45 \\
   G & $\equiv$0\farcs00 & $\equiv$0\farcs00\\ 
  \hline
\end{tabular}
\caption{Astrometry of the lensed images of SDSSJ1402.
\label{imtable}}
\tablenotetext{}{(Positions are determined by
eye from the F435W residual
image; we adopt conservative $0 \farcs 1$ errors in right ascention
and declination for the lens modeling described in \S~\ref{lensmod}.)}
\end{table}

\subsection{Gemini GMOS-N IFU}

\begin{figure}
\begin{center}
\scalebox{1.2}{\plotone{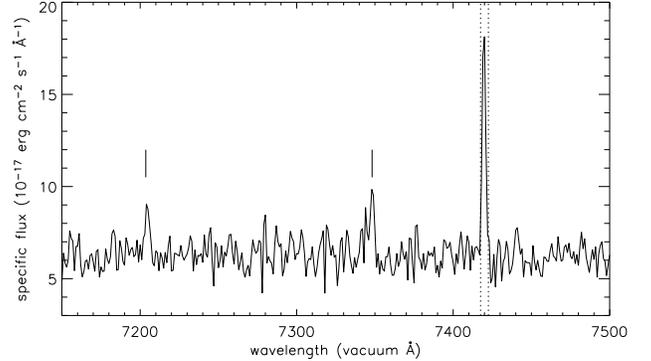}}
\end{center}
\caption{Summed GMOS-N
IFU fiber spectra for fibers near the narrowband image
positions seen in Figure~\ref{ims}
(without cotinuum subtraction).
Redshifted H$\beta$ 4863~\AA\ and [OIII] 4960~\AA\ are
detected at the expected positions, as shown by tick marks above the
spectrum.  Dotted lines show the narrowband-image wavelength range
used to generate the IFU image of Figure~\ref{ims}.
\label{ifu_fig}}
\end{figure}

To solidify the lensing hypothesis over alternative
explanations for the observed features, we look to
spatially resolved spectroscopy of SDSSJ1402 obtained with the
integral-field unit (IFU) of the Gemini Multi-Object Spectrograph
(GMOS-N) on the 8-meter Gemini North telescope at Mauna Kea
\citep{hook_gmos, jas_gmos_ifu}.
The observations were made using the R600 grating and a
Sloan $i$-band blocking filter to prevent spectral
overlap and allow use of the full $5 \arcsec \times 7 \arcsec$
IFU field of view.  The total
integration time was 3$\times$900 seconds,
and the data were reduced using custom IFU software
written in IDL (Bolton \& Burles, in preparation).
In addition to the ACS data,
Figure~\ref{ims} also shows a reconstructed IFU
narrow-band
image of a 5-\AA\ window about the redshifted
[O III] 5008.24\AA\ emission,
after subtraction of a linear fit to the LRG continuum at that
wavelength.  The emission is spatially coincident with the assumed
lensed images in the HST-ACS data.  Independent confirmation of the
identification of this line emission as
redshifted [OIII] is provided by summing the spectra in
circular apertures about the appoximate image positions to obtain the
spectrum presented in Figure~\ref{ifu_fig}, showing both lines of the
[OIII] doublet in the expected 1:3 ratio, as well as H$\beta$.
We thus confirm that the redshift of the faint images is much
higher than that of the LRG, further supporting a lensing
interpretation. \\

\section{Gravitational-Lens Model}
\label{lensmod}

Adopting a lensing hypothesis for SDSSJ1402,
we model the lens as a simple singular isothermal ellipsoidal (SIE)
mass model without external shear, using the astrometry of the four
images as constraints. Using the mass normalization from
\citet{kormann_sie}, the best-fit ($\chi^2=2.1$ for NDF=3) gives
an Einstein radius $b=1\farcs35\pm 0\farcs05$~(68\% CL), corresponding
to a SIE velocity dispersion of $\sigma_{\rm SIE} = 292 \pm
6$~km/s\footnote{Recall that the luminosity-weighted stellar
dispersion from the SDSS is $\sigma_{*}^{\rm sloan}=267 \pm
17$~km/s (within a 3$''$ diameter aperture). {\sl Prior} to the ACS
observations we predicted $b=1 \farcs 12 \pm 0\farcs 14$
based on this dispersion,
leading to the selection as lens candidate.}
and a physical radius of $(4.9 \pm 0.2)h_{65}^{-1}$ kpc.
The axial ratio and
position angle (in the frame of Figure~\ref{ims}) of the
ellipsoidal equal surface density contours are
$q=0.78_{-0.11}^{+0.12}$ and $\theta = 63.6^{+2.2}_{-2.2}$~degrees,
respectively.
This lens-model ellipticity
is in good agreement with the
ellipticity of the galaxy
isophotes given in \S~\ref{acs}.
The galaxy and lens-model position angles are
also significantly aligned, although
they differ formally by $\approx 3 \sigma$.

The total mass-to-light ratio within the
Einstein radius is robustly determined by the data,
and we compute it as follows.
The total mass enclosed by the critical curve is $(30.9
\pm 2.3)\times 10^{10}h_{65}^{-1}$~M$_\odot$: we derive this
from the fitted lens model, but it is in fact
largely model independent.
From aperture photometry,
the corresponding enclosed magnitude
is $B_{435} = 20.28$.
Using a {\tt specBS} template fitted to the
spectrum of SDSSJ1402 and an ACS-F435W filter curve,
we compute a $k$-correction of 1.20 for the lens galaxy.
Combining this with a
distance modulus of 40.17 from our assumed cosmology, we obtain
an absolute $B_{435}$ magnitude of $-21.09 \pm 0.05$
enclosed by the lensed images, with an estimated
0.05-magnitude systematic error in the conversion.
Finally, we convert from the $AB$ system to Johnson $B$
via $B_{435} = B - 0.11$, giving
an absolute $B$ magnitude of $-20.98 \pm 0.05$.
For an absolute solar magnitude of
$M_{B,\odot} = 5.47$ \citep{aaq_2000},
the rest-frame $B$-band
luminosity within the Einstein radius is then
$L_{B,{\rm encl}} = (3.8 \pm 0.2) \times 10^{10}h_{65}^{-2} L_{B,\odot}$.
Thus the enclosed $B$-band mass-to-light ratio is
$\Upsilon_{B} = (8.1 \pm 0.7)h_{65}$ in solar units.

If
we assume the stars to be test particles embedded in an
isothermal mass distribution with the above given mass,
with a
luminosity density derived from the stellar surface brightness (e.g.\
see KT03 or TK04 for details), we find a luminosity-weighted stellar
dispersion of $\sigma_{*}(<3 \arcsec) \approx 270$~km\,s$^{-1}$
within the
SDSS apperture, assuming an isotropic dispersion tensor.
This agrees very well with the observed stellar
dispersion.
Future work will present a more detailed dynamical analysis
based on the determination of an extended kinematic
profile or field (e.g.\ with integral-field spectroscopy).

\begin{figure}[t]
\begin{center}
\scalebox{1.1}{\plotone{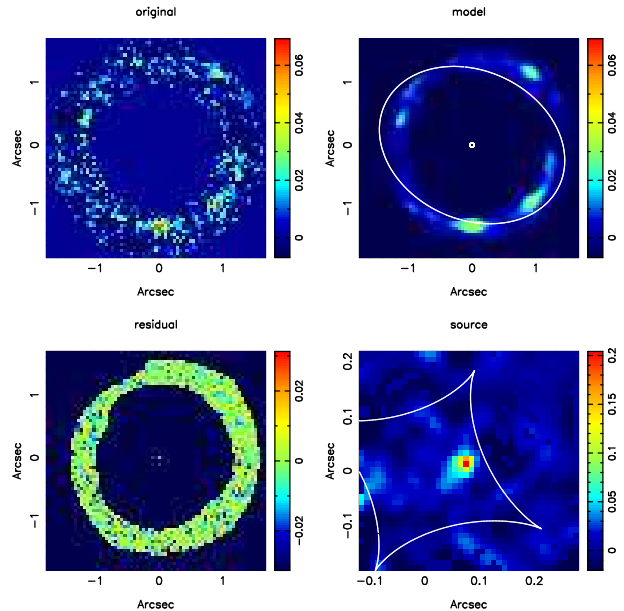}}
\end{center}
\caption{Gravitational-lens inversion of SDSSJ1402 in the
F435W band.  Upper left panel shows
the galaxy-subtracted HST-ACS image within a suitable
annulus.  The best model of the system (upper right) is the mapping of
the source model (lower right) on to the image plane using the
best-fit SIE mass model. The critical curve and caustic are plotted
on the image and source models, respectively.  Data $-$ model residual
images are shown in the lower left panel.
\label{model_comb}}
\end{figure}

With this mass model, we proceed to {\sl invert} the lensed images to
obtain the structure of the source
(e.g.\ \citealt{warren_dye_03}; see also \citealt{tk04}
for more details).
We regularize the solution by
adding a term to the penalty function that suppresses the curvature
(see Press et al. 1992) of the resulting source brightness
distribution. During the single-step inversion process the mass model
remains fixed. We use PSF models generated with Tiny-Tim, although the
results are insensitive to the precise PSF model because of the low
S/N of the images.
The F435W results are shown in Figure~\ref{model_comb},
for a regularization that gives
$\chi^2/{\rm NDF}\approx1.0$.  We
see
that the source has a
roundish monotonically-decreasing brightness distribution,
close to and inside the fold caustic.
In the F814W filter (not shown), the source is found at the
same position and with similar structure. Also shown are the lensed
source model
and residuals. The latter show no
significant remaining structure and are distributed Gaussian.
We find that models
with parameters
within the 68\% CL of the
best-fit SIE model values
give
similar results, but can not be distinguished at a significant level
because of the low S/N of the images. Hence, the lens inversion should
for now be regarded only as a consistency check of the lens model.

\section{Conclusions}

Based on the following
evidence, we conclude that SDSSJ1402 is a
lens system:

\begin{enumerate}
\item The discovery of emission-lines in the SDSS spectrum at a redshift
 significantly larger than that of the high-$\sigma_{*}$
 LRG and within a 1.5$''$ radius from the galaxy centroid,
 and confirmed by follow-up GMOS-N IFU spectroscopy.
\item The discovery of four images (with two
 tangentially stretched) in a typical quad-lens-like configuration
 around the galaxy in {\sl Hubble Space Telescope}
 images, with similar colors and positions between F814W and
 F435W filters.
\item The spatial coincidence of these images with the higher-redshift
 line emission, as shown by GMOS-N IFU spectroscopy.
\item The excellent goodness-of-fit ($\chi^2/{\rm NDF}\approx 0.7$)
 for the SIE mass model (no external shear).
 The stellar velocity dispersion predicted by this mass model
 and the observed luminosity density is in excellent agreement
 with the SDSS value,
 and the galaxy and lens-model ellipticities
 and position angles are in significant agreement
 with one another. 
\item A direct inversion of the lensed image---based solely on the
  SIE mass model determined from the image positions---leads to a
  very simple compact source structure in both F814W and F435W filters
  at the same position
  near
  the inside of the
  fold caustic.
\end{enumerate}

Hence, despite the relatively low signal-to-noise of the lensed
images, the spectral and imaging data of SDSSJ1402 presented in this
paper, in conjunction with the self-consistent lens model,
convincingly show that
the {\sl first} candidate observed
by the SLACS Survey is also
the {\sl first} genuine lens system from the program!
The lens geometry and HST photometry yield a (largely
model-independent) measurement of 
$(30.9 \pm 2.3)\times 10^{10}h_{65}^{-1}$~M$_\odot$
and a rest-frame $B$-band mass-to-light ratio of
$(8.1 \pm 0.7)h_{65}$ times solar
within the $(4.9 \pm 0.2)h_{65}^{-1}$-kpc cylinder enclosed by the
critical curve.
Spatially resolved kinematic observations
of this system will permit more detailed lens/dynamical
modeling, which will constrain the radial mass profile and
the relative fraction of luminous to dark matter
in the central region of the lensing galaxy.\\

\acknowledgments

TT acknowledges support from NASA through Hubble Fellowship grant
HF-011167.01.

L.A.M. acknowledges support from the 
Spitzer Legacy Science Program, provided by NASA through contract 
1224666 issued by the Jet Propulsion Laboratory, California Institute of 
Technology, under NASA contract 1407.

The authors thank Hsiao-Wen Chen for the
gracious provision of an ACS-F435W filter curve.
We are grateful for the scheduling work done by Galina Soutchkova, the 
Program Coordinator for this HST program (SNAP-10174).


\end{document}